# Realizing stable zig-zag polymeric nitrogen chains in P-N compounds

Chengfeng Zhang[1,2], Guo Chen[1,2], Yanfeng Zhang[1,3], Jie Zhang[1], and Xianlong Wang[1,2]*

[1] *Key Laboratory of Materials Physics, Institute of Solid State Physics, HFIPS, Chinese Academy of Sciences, Hefei 230031, China*

[2] *University of Science and Technology of China, Hefei 230026, China*

[3] *Institutes of Physical Science and Information Technology, Anhui University, Hefei 230601, China*

———

*Author to whom all correspondence should be addressed: xlwang@theory.issp.ac.cn

## ABSTRACT

The zig-zag Nitrogen (N) chain similar to the Ch-N structure has long been considered a potential high energy density structure. However, all previously predicted zig-zag N chain structures similar to Ch-N exhibit imaginary frequencies in their phonon spectra at 0 GPa. Here, we conducted a systematic investigation of P-N compounds using first-principles calculations, uncovering a series of structurally similar stable phases, C2/m-$PN_x$ (x = 6, 8, 10, 12, 14), in which N forms zig-zag N chains similar to those in Ch-N. In P-N compounds, the longest zig-zag N chain that can theoretically remain stable under ambient pressure is the N chain composed of 14 N atoms in C2/m-$PN_{14}$. If the N chain continues to grow, inter-chain vibrational imaginary frequencies will arise in the system. Notably, N chains with an even number of atoms are more likely to be energetically favorable. The five C2/m-$PN_x$ phases and one metastable phase (R-$PN_6$) exhibit both remarkable stability and excellent detonability at ambient pressure, positioning them as promising candidates for high-energy-density materials. In addition, the R-$PN_6$ is the first structure to stabilize the $N_6$ ring through covalent bonding, with the covalent network contributing to its high hardness (47.59 GPa).

# INTRODUCTION

Polymeric N materials have garnered significant attention due to their cleanness and high energy density, which has important potential applications as High-energy-density materials (HEDMs). Theoretical studies have reported a variety of novel polymeric N structures, including typical molecular crystals like $N_{16}$ and $N_{10}$ [1,2], cage-like $N_{10}$ [3], chain-like phases such as Ch-N and Cmcm-N [4,5], layered structures such as BP, PP, LB, ZS, and A7 phase [6-9]. Network formations including cg-N and Pnnm phases [10,11]. In 2004, cg-N was successfully synthesized under extreme conditions of 2000 K and 110 GPa [10], marking the first experimental synthesis of polymeric N structures predicted by theory. Subsequently, BP-N, LP-N, and HLP-N were also successfully synthesized under the high pressure of 140 GPa, 150 GPa, and 250 GPa, respectively [12-14]. In addition to relying on high temperature and pressure, recent years have seen the successful synthesis of polymerized N through chemical and plasma methods, providing new perspectives for its synthesis [15-17].

To obtain more polymeric N under mild conditions, N-rich materials doped with various elements gained lots of attention. Similar to pure N materials, N-rich compounds also exhibit a variety of N configurations. The configurations of N in these materials, as reported in theoretical studies, primarily include zero-dimensional ring structures [18-20], one-dimensional chain structures [21-26], two-dimensional layered structures [27,28], and three-dimensional network structures [29,30]. In zero-dimensional ring structures, The $N_5$ ring tends to acquire electrons from impurity atoms, thus achieving good stability [31]. One-dimensional chain structures, on the other hand, generally require lower synthesis pressures [32]. As a result, many experimental synthesis of zero-dimensional [33-36] and one-dimensional structures [37-40] are reported. Interestingly, the one-dimensional zig-zag N chain, which has been reported in pure N systems under high pressure, has yet to be reported in the theoretical or experimental studies of N-rich materials. This zig-zag N chain, composed of individual N dimers, is believed to possess a rich variety of physical properties, including superconductivity and high energy density [5]. Therefore, it is very urgent to find the standard zig-zag N chain in N-rich systems. Previous research has indicated that the introduction of P into pure N materials can promote the polymerization of N [41]. Furthermore, the electronic structure of P is similar to that of N, and doping P may successfully induce the zig-zag N chain reported in pure N systems within N-rich systems. Therefore, we have decided to conduct in-depth research on P-N compounds.

In this work, structural searches and first-principles calculations were employed. C2/m-$PN_x$ (x = 6, 8, 10, 12, 14) containing zig-zag N chains were discovered, and these structures were found to be dynamically stable under ambient pressure, exhibiting good detonation performance. The reason why infinitely long zig-zag N chains, such as those in Ch-N structures, cannot be synthesized experimentally is illustrated. In addition, we have discovered R-$PN_6$ containing an $N_6$ ring, and this zero-dimensional cyclic structure stabilized by covalent bonds is reported for the first time in N-rich compounds.

## METHODS

The structure prediction for the P–N system was performed using the particle swarm optimization (PSO) methodology implemented in the CALYPSO code [42]. Structural searches were conducted on $PN_x$ (x = 3, 4, 5, 6, 7, 8, 10, 12, 14) compounds across pressures ranging from 0 to 150 GPa with 1-3 formula units. First-principles calculations of the structure and electronic properties based on the density functional theory, utilizing a plane-wave basis set, were performed using the Vienna ab initio simulation package (VASP) [43]. The generalized gradient approximation (GGA) [44,45] parameterized by the Perdew-Burke-Ernzerhof (PBE) [46] exchange-correlation functional, along with the projected augmented wave (PAW) [47] was employed for the calculations. Computational accuracy was ensured with a grid density of 0.03 Å$^{-1}$ and an energy cutoff of 520 eV. The formation enthalpies $\Delta H_f$ of $PN_x$ compounds was determined using the formula $\Delta H_f(PN_x) = [H(PN_x) - H(p) - xH(N)] / (1 + x)$. Where $H(PN_x)$ is enthalpy of the $PN_x$ compound. $H(P)$ and $H(N)$ are the enthalpies per atom of Pm$\bar{3}$m-P, ε-N, and cg-N. The phonon dispersion relations were calculated using the Phonopy code based on the finite displacement method [48]. The PBE-vdW method was employed for all structures [49]. Detonation velocity and detonation pressure were derived using the Kalet-Jacobs equation [50], COHP was calculated using the LOBSTER program [51]. The Bader charge of the structure is calculated using VTST program [52-55]. Additionally, VASPKIT is also adopted in our work [56].

## RESULTS AND DISCUSSION

The convex hull curve is constructed based on the formation enthalpies of thermodynamically stable structures, and the convex hull curves at different pressures are shown in the Fig. 1 (a). We can find that at 0 GPa, only $P_3N_5$ exits on the convex hull, which consists with previous experimental observations and theoretical simulations [57,58]. With pressure increasing to 43 GPa, $PN_6$ with C2/m symmetry become stable on the convex hull. Furthermore, we can notice that other the $PN_x$ (x = 4, 8, 10, 12, 14) with the same symmetry of C2/m are also very close to the convex hull. For example, $PN_8$ and $PN_{10}$ are only 27 meV/atom and 65 meV/atom above the convex hull curve, respectively. When the pressure is increased to 53 GPa, besides C2/m-$PN_6$, C2/m-$PN_x$ (x = 8, 10, 12, 14) also became stable, and corresponding structures are shown in Fig. 2 (a-e). Recently, theoretical structure searching was applied on the compositions of $PN_x$ (x = 2, 3, 4), they found that $PN_4$ will became stable at pressure of 50 GPa [58]. However, if components with higher N content are included, the previous reported P2/m-$PN_2$, Immm-$PN_3$, and C2/m-$PN_4$ is located above the convex hull. The C2/m-$PN_6$ continues to remain energetically favorable as the pressure reaches 100 GPa, where C2/m-$PN_4$ also reappears in our search. The stable pressure range of C2/m-$PN_x$ is shown in Fig. 1 (b), and C2/m-$PN_6$ exhibits the widest stable pressure range from 43 GPa to 100 GPa, indicating that C2/m-$PN_6$ possesses the highest stability among the N-rich phases in our predictions. C2/m-$PN_8$ becomes metastable at approximately 58 GPa, while C2/m-$PN_{10}$ transitions to a metastable phase at around 55 GPa. Similarly, both C2/m-$PN_{12}$ and C2/m-$PN_{14}$ undergo a transition to metastable phases at approximately 58 GPa.

As shown in Fig. 2 (a-e), all of newly discovered N-rich phases, C2/m-$PN_x$ (x = 6, 8, 10,

12, 14), exhibit striking structure similarities, where N atoms forming zig-zag chains. As the N content increases, the length of the zig-zag N chains also increases. In these stable phases, the coordination number of P is six, and six N atoms form an octahedron around the central P atom. The main difference between these phases reflected in the variable number of N atoms within their chains. Furthermore, the local coordination environment of P in C2/m-$PN_x$ (x=6, 8, 10, 12, 14) is consistent with that in $P_3N_5$ under high pressure, with a coordination number of six (see Fig. S1), whereas under low pressure, the coordination number of P in α-$P_3N_5$ is four. This suggests that the formation of P-N octahedra in the P-N system under high pressure is more favorable for the stability of the system. Under high pressure conditions, as the N content increases, the P-N octahedra will act as a framework, stabilizing the zig-zag N chains formed by the newly added N atoms.

For R-$PN_6$, the space group is R$\bar{3}$m, and the coordination number of the P atoms is six. Unlike C2/m-$PN_x$, the N aggregation in R-$PN_6$ takes the form of a non-coplanar $N_6$ ring. Due to the small electronegativity difference between P and N, the six P atoms surrounding the $N_6$ ring form six covalent bonds with each N atom in the $N_6$ ring. These six covalent bonds collectively stabilize the central $N_6$ ring. In prior N-rich environments, $N_6$ rings were stabilized by ionic bonds with dopant atoms [19,59-61]. The $N_6$ ring in the R-$PN_6$, stabilized through covalent bonding with dopant elements, is reported by us for the first time. Although R-$PN_6$ is a metastable phase, when the pressure increases to 100 GPa, R-$PN_6$ is located just 66 meV/atom above the convex hull, indicating that this structure has potential for synthesis under high temperature and high pressure [3,13]. The discovery of R-$PN_6$ provides new insights for finding N-rich compounds containing the $N_6$ ring.

At high pressure, the phonon dispersion curves of both the thermodynamically stable phase C2/m-$PN_x$ (x=6, 8,10,12,14) and the metastable phase R-$PN_6$ show no imaginary frequencies, indicating their dynamical stability under high pressure (Fig. S2). When the pressure is reduced to 0 GPa, the six phases remain dynamically stable, indicating that they have the potential to be quenched to ambient pressure (Fig. 3 (a)-(c) and Fig. S3). Furthermore, the elastic constant matrices of these six new phases under ambient conditions meet the criteria for mechanical stability (Table S1-6). As indicated by the green-marked regions in Fig. 3 (a)-(c), with the increase in N content, phonon softening occurs along the high-symmetry path A-Γ in C2/m-PN$_x$. Here, we refer to the phonon vibration mode with the lowest frequency at the high-symmetry point A of C2/m-PN$_x$ as Mode A. The frequency of Mode A in C2/m-PN$_x$ (x = 6, 8, 10, 12, 14, 16) is shown in Fig. 3 (d). As the N content increases, the frequency of Mode A continuously decreases. In C2/m-PN$_{14}$, the frequency of Mode A drops to 0.7 THz. This phonon frequency, which is close to 0, suggests that the system has reached the limit of dynamic stability. When the N content is further increased to C2/m-PN$_{16}$, the frequency of Mode A becomes negative, leading to the instability of C2/m-PN$_{16}$ and systems with higher N content (The structure and phonon dispersion curves can be found in Fig. S4).

To elucidate the deep mechanism behind the phonon softening in the C2/m-PN$_x$ system induced by the increase in N content, we conducted further studies on the phonon vibration modes. Fig. 3 (e) and Fig. 3 (f) illustrate the vibrational mode A in the C2/m-PN$_6$ and C2/m-PN$_8$, where the P-N octahedron, due to its greater mass, primarily governs the oscillations of the zig-zag N chain as a whole. Notably, the same phonon vibrational mode as mode A is observed in the C2/m-PN$_x$ systems (x = 6, 8, 10, 12, 14, 16). As the N chain lengthens, its mass increases, and the P-N octahedron becomes progressively less effective in restraining the zig-zag N chain, resulting in a reduction in the frequency of mode A, which ultimately leads to the destabilization of the system. Therefore, in the C2/m-PN$_x$ system, the length of the N chain must have a limit. Through the phonon dispersion curves of the C2/m-PN$_x$, it is concluded that the longest zig-zag N chain capable of maintaining dynamical stability in the system consists of 14 N atoms (Fig. 3 (a)-(c), Fig S4).

In order to study the variation of N-N bond lengths in N chains composed of different N atoms in the C2/m-PN$_x$ system at 0 GPa, we present Fig. 4. The dashed line in the figure indicates the N-N bond length in Ch-N. It can be seen from this line that each N-N bond in Ch-N has a length of 1.30 Å. Regardless of the length of the zig-zag N chain in the C2/m-PN$_x$ system, the N-N bond lengths at both ends of the zig-zag N chains remain consistent overall due to their close proximity to the P-N octahedra. Furthermore, the lengths of the N-N bonds oscillate from both ends toward the center, with the amplitude of oscillation decreasing as they approach the center. In C2/m-PN$_{14}$, due to the relatively short length of the N chain, the P-N octahedra at both ends can exert a significant influence on the entire N chain. As a result, the distribution of bond lengths in the zig-zag N chain of $PN_{14}$ differs considerably from that in the zig-zag N chain of Ch-N (Fig. 4 (a)). As the length of N chain increases, in $PN_{16}$ and $PN_{18}$, the oscillation of N-N bonds around the central points of decreases, the N-N bond lengths in the central region of the chain will approach that of Ch-N (1.3 Å) (the darker areas in Fig. 4 (b) and Fig. 4 (c)). For example, in $PN_{18}$, the two central N-N bond lengths are 1.29 Å and 1.30 Å, results in the instability of $C2/m$-$PN_{16}$ and other structures with longer zig-zag N chains under ambient pressure, similar to the behavior observed in Ch-N structures (Fig S5). In summary, the alternating bond length variation of the N-N bonds is beneficial for the stability of the zig-zag N chain.

Interestingly, in the $PN_x$ system, no stable structures with an odd number of N atoms have been found. Therefore, in the $PN_x$ system, structures with an even number of N atoms tend to be more energetically favorable than those with an odd number of N atoms. To investigate why the $PN_x$ system is more energetically favorable when number of N atoms is even, the Electron Localization Function (ELF) for the energetically favorable phase $C2/m$-$PN_x$ (x=6, 8, 10, 12, 14) is calculated. As shown in Fig. 5, there is significant electron localization between N atoms, indicating the formation of strong covalent bonds, which constitute the backbone of the zig-zag chain. Moreover, lone pair electrons on both sides of the N chain aligned linearly alone the chain (The regions in the Fig. 5 where the ELF value is close to 1), effectively maintained by the zig-zag N chain structure of $C2/m$-$PN_x$, which minimizes the system's Coulomb repulsion. According to Valence Shell Electron Pair Repulsion (VSEPR) theory [62], a system with smaller Coulomb repulsion implies better stability. When the number of N atoms is even, the lone pair electrons on both sides of the zig-zag chain are evenly distributed, which minimizes the total Coulomb repulsion among them. For systems with an odd number of N atoms, it is impossible to form equal amounts of lone pair electrons on both sides of the zig-zag N chain, so the Coulomb repulsion in odd numbered systems is higher than in even numbered systems. As a result, in the P-N compounds, structures with an even number of N atoms tend to be more thermodynamically stable than those with an odd number of N atoms.

Fig. 6 (a) shows the electronic density of states (DOS) of $C2/m$-$PN_{14}$ at 0 GPa. The DOS

shows that the system does not exhibit a band gap, indicating that C2/m-PN$_{14}$ displays metallic properties. The DOS near the Fermi level is mainly contributed by N_2p electrons, and thus the metallicity of the system primarily originates from the N_2p electrons. The Bader charge (Fig. S6) indicates that in C2/m-PN$_{14}$, the P atoms transfer 2.8 electrons to the N atoms. These transferred electrons can effectively facilitate the polymerization of N atoms [63].

To further explore the bonding characteristics, we calculated the partial crystal orbital Hamilton population (pCOHP) between P and neighboring N atoms in C2/m-PN$_{14}$ at 0 GPa, as well as between N atoms in the N chain, which is depicted in Fig. 6 (b). Below the Fermi level, bonding states of P-N are fully occupied, while antibonding states of P-N above the Fermi level remain completely unoccupied, indicating the stability of the P-N bond. For the N-N bond, bonding states are fully occupied, and some antibonding states are partially occupied. The ICOHP for the five stable phases was calculated as depicted in Fig. 6 (c). Although some antibonding orbitals of N are occupied, the ICOHP between N-N remains less than zero, indicating that the N-N bond is stable. As the N chain grows, the bond strength of the P-N bond gradually decreases, while the bond strength of the N-N bond gradually increases. When the N chain reaches N$_{10}$, the strength of the N-N bond surpasses that of the P-N bond. To explain the cause of the change in bond strength of the P-N and N-N bonds with N chain length in C2/m-PN$_x$, the bond lengths of the P-N and N-N bonds in the system are shown in Fig. 6 (d). From Fig. 6 (d), it can be observed that as the length of the N chain increases, the average bond length of the P-N bond increases, whereas the average bond length of the N-N bond decreases. As a result, the increase in N content will cause the P-N bond length to increase and the N-N bond length to decrease, ultimately leading to a decrease in the bond strength of the P-N bond and an increase in the bond strength of the N-N bond.

Energy density (E$_d$), detonation velocity (V$_d$), and detonation pressure (P$_d$) are crucial parameters for assessing the performance of HEDMs. We calculated the energy density of PN$_x$ compounds under ambient pressure using the dissociation pathway 6PN$_x$→2P$_3$N$_5$ + (3x-5)N$_2$. α'-P$_3$N$_5$ [64] and α-N$_2$ [65] were chosen as the ground state phase. Subsequently, P$_d$ and V$_d$ were estimated using the Kamlet-Jacobs equation [50]. As depicted in Fig. 7, these five stable phases demonstrate significantly higher $P_d$, $V_d$, and volumetric energy density ($E_v$) compared to traditional HEDMs such as TNT and HMX [25,66]. The metastable phase R-PN$_6$ exhibits superior $P_d$, $V_d$, $E_v$, and $E_d$ compared to TNT, HMX, and the five stable phases. In particular, the $E_d$, $E_v$, $V_d$, and $P_d$ values for R-PN$_6$ are 5.915 KJ/g, 22.087 KJ/cm$^3$, 11.447 Km/s, and 81.415 GPa, respectively. Additionally, R-PN$_6$ possesses a Vickers hardness of 47.59 GPa, positioning it as a promising superhard material.

## CONCLUSION

In this paper, we have realized zig-zag polymeric N chains, similar to those found in Ch-N, within C2/m-PN$_x$ (x=6, 8, 10, 12, 14), which can be quenched to ambient pressure and exhibit excellent detonation performance. Among them, the longest zig-zag N chain is found in C2/m-PN$_{14}$, composed of 14 N atoms. By analyzing the phonon properties of C2/m-PN$_x$ (x = 6, 8, 10, 12, 14, 16, 18), it was found that with the increase in N chain length, the P-N octahedra in the system struggle to maintain the stability of the zig-zag N chain, resulting in phonon softening and ultimately causing structural instability. Electronic structure analysis shows that zig-zag N chains composed of an even number of N atoms form equal amounts of lone pair electrons on both sides of the chain, effectively reducing electron Coulomb repulsion. As a result, zig-zag N chains with an even number of N atoms tend to be more energetically favorable. The metastable phase R-PN$_6$ contains an N$_6$ ring, which is stabilized for the first time by covalent bonds. Its high energy density (5.915 kJ/g), high Vickers hardness (47.59 GPa), and kinetic stability under ambient pressure suggest that it has potential as HEDMs and superhard materials. This study provides theoretical insights for the development of novel HEDMs.

## ACKNOWLEDGEMENTS

This work is supported by the CASHIPS Director's Fund (Grant No. YZJJ202207-CX). The calculations were partially performed at the Center for Computational Science of CASHIPS, the ScGrid of the Supercomputing Center, and the Computer Network Information Center of the Chinese Academy of Sciences, as well as at the Hefei Advanced Computing Center.

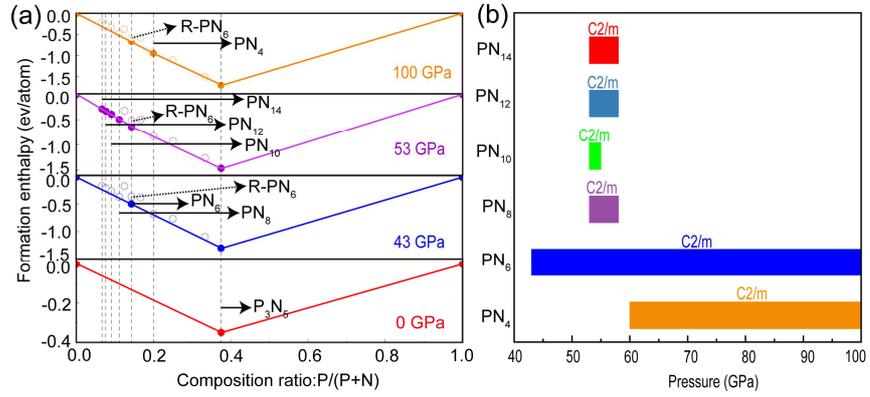

**FIG. 1.** The thermodynamic stability of P-N compounds. (a) Formation enthalpies of vaious $PN_x$ (x=2, 3, 4, 5, 6, 7, 8, 10, 12, 14) compounds (with respect to pure P and pure N). Stable phases are represented by solid dots, while unstable/metastable phases are represented by hollow dots. The hollow circle corresponding to $PN_6$ is R-PN6, which is just located 66 meV/atom above the convex hull curve at 100 GPa. (b) Pressure-composition phase diagram of the predicted P-N phases.

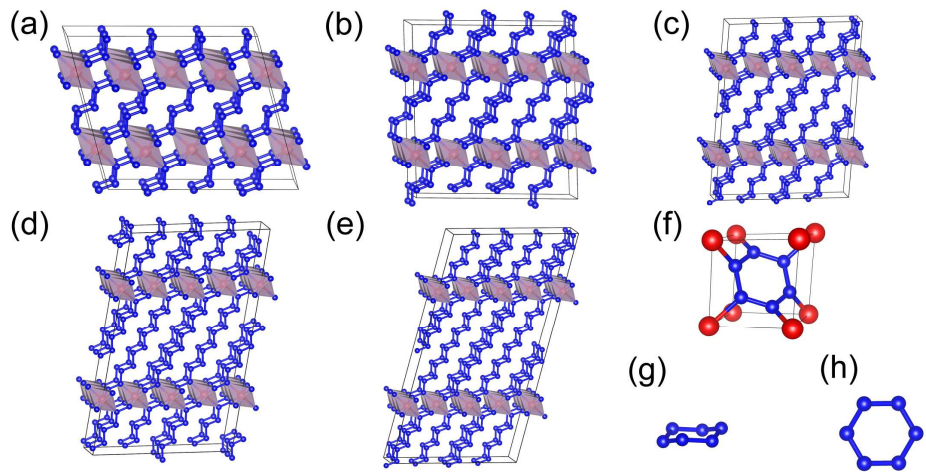

**FIG. 2.** Crystal structures of P-N compounds at 0 GPa. (a) C2/m-PN$_6$, (b) C2/m-PN$_8$, (c) C2/m-PN$_{10}$, (d) C2/m-PN$_{12}$, (e) C2/m-PN$_{14}$, (f) R-PN$_6.$ (g) Side perspective of the N$_6$ ring in R-PN$_6$ .(h) Top view of the N$_6$ ring in R-PN$_6$. Red represents P atoms, and blue represents N atoms.

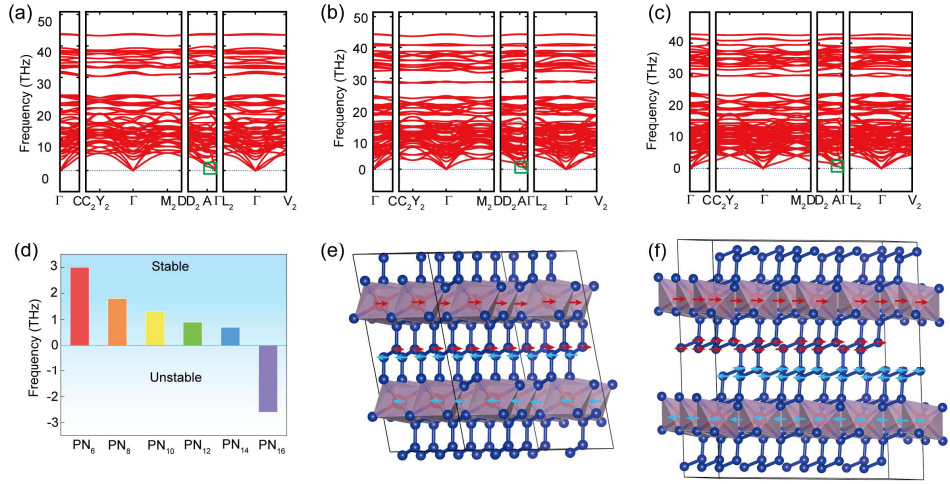

**FIG. 3.** Phonon properties of C2/m-PN$_x$. (a)-(c) show the phonon dispersion curves of C2/m-PN$_{10}$, C2/m-PN$_{12}$, and C2/m-PN$_{14}$ at 0 GPa. The green-highlighted sections in the phonon dispersion curves indicate regions where phonon softening occurs as the length of N chain increases. (d) The frequency of the lowest-frequency phonon vibration mode (Mode A) at the high-symmetry point A in the C2/m-PN$_x$ (x=6, 8, 10, 12, 14, 16). (e) and (f) show Mode A in the C2/m-PN$_6$ and C2/m-PN$_8$. The direction of the arrow represents the direction of the vibration.

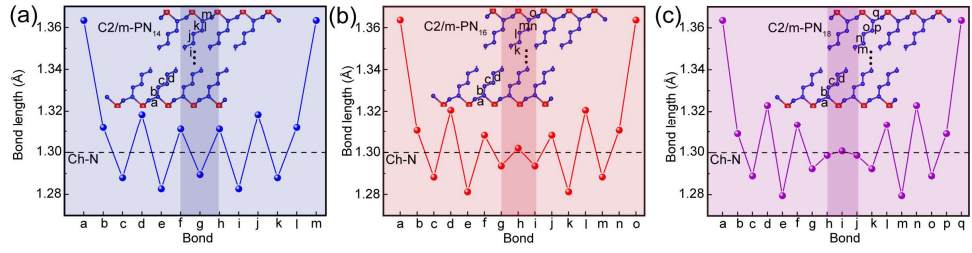

FIG. 4. N-N bond length of Ch-N and C2/m-PN$_x$ (x=14, 16, 18) at 0 pressure. The letters on the x-axis in the figure represent different N-N bonds in C2/m-PN$_x$ (x=14, 16, 18). (a)-(c) represent C2/m-PN$_{14}$, C2/m-PN$_{16}$, and C2/m-PN$_{18}$, respectively. The dashed lines represent the N-N bond lengths in the Ch-N structure. The dark areas in the figure represent the N-N bonds in the center of the N chains in the C2/m-PN$_x$ system.

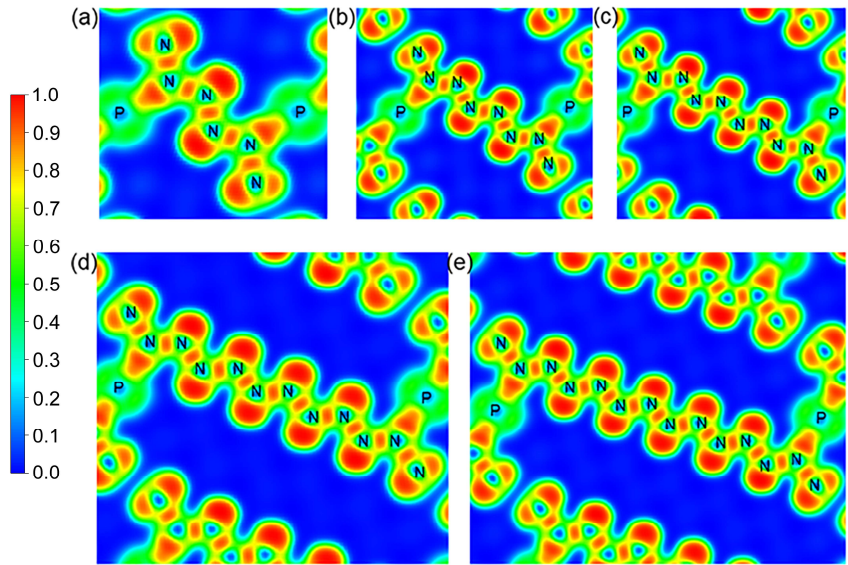

**FIG. 5.** ELF for C2/m-PN$_x$ at 0 GPa. The ELF values on both sides of the N chain are higher, with clear electron localization corresponding to the lone pair electrons in the system. (a) C2/m-PN$_6$, (b) C2/m-PN$_8$, (c) C2/m-PN$_{10}$, (d) C2/m-PN$_{12}$, and (e) C2/m-PN$_{14}$.

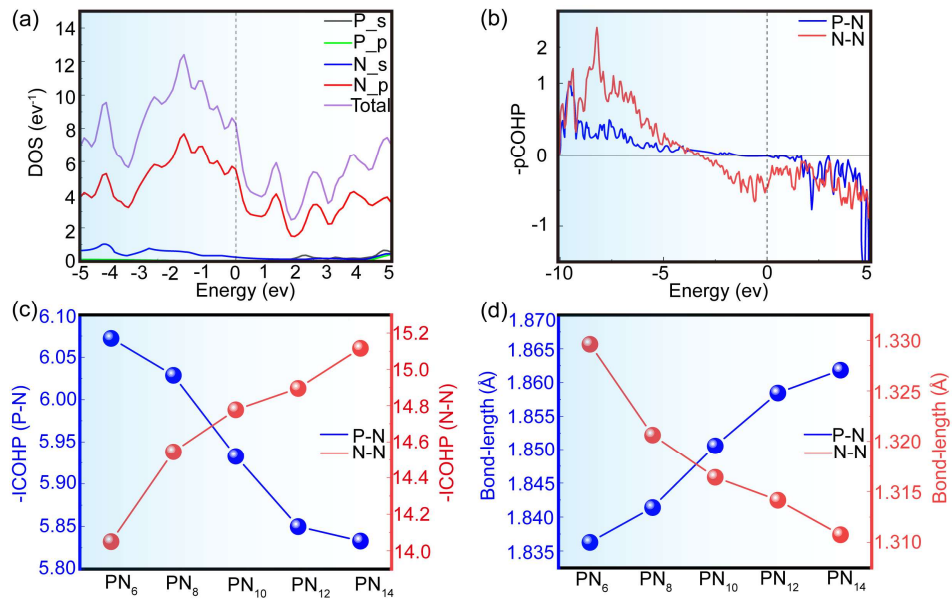

**FIG. 6.** (a) PDOS of C2/m-PN$_{14}$ at 0 GPa, (b) pCOHP of P-N bonds and N-N bonds in C2/m-PN$_{14}$ at 0 GPa, (c) ICOHP of P-N and N-N bonds in stable C2/m-PN$_x$ at 0 GPa, (d) Average bond length of P-N and N-N bonds in stable C2/m-PN$_x$ at 0 GPa.

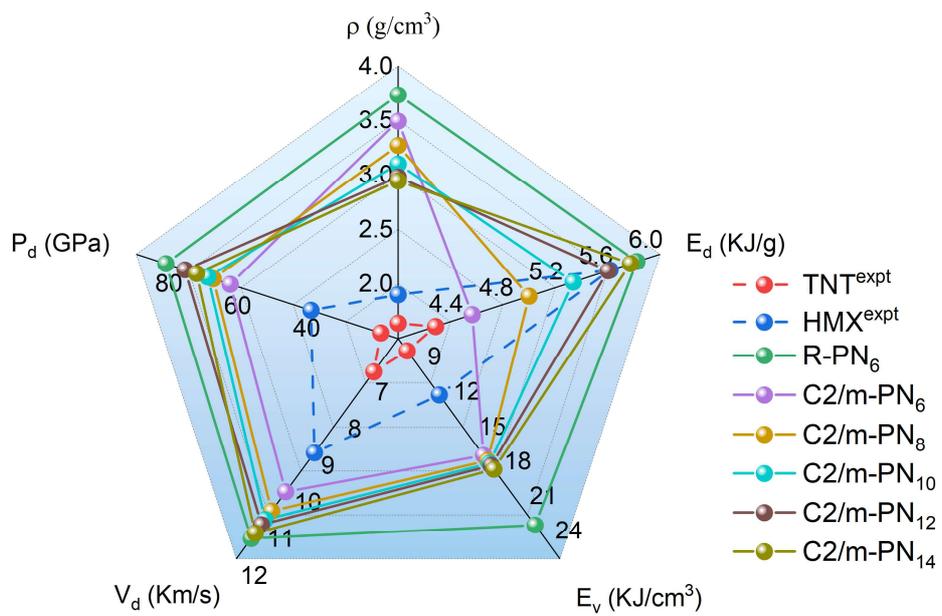

**FIG. 7.** Calculated energy density ($E_d$), volumetric energy density ($E_v$), detonation velocity ($V_d$), detonation pressure ($P_d$), and density ($\rho$) of P-N compounds.